\documentclass[aps,pra,twocolumn,superscriptaddress,nobalancelastpage]{revtex4-1}
\usepackage{amsmath,amssymb,mathtools}
\usepackage[dvipdfmx]{graphicx}
\usepackage[colorlinks=true,bookmarks=true,citecolor=blue,linkcolor=blue,urlcolor=blue,breaklinks=true]{hyperref}

\graphicspath{{fig/}}
\usepackage{dcolumn}
\usepackage{bm}
\usepackage[T1]{fontenc}
\usepackage{textcomp}
\usepackage{color}
\usepackage{mathrsfs}

\begin{document}

\title{Violation of the magnonic Wiedemann-Franz law in the strong nonlinear regime
}

\author{Kouki Nakata}
\affiliation{Advanced Science Research Center, Japan Atomic Energy Agency, Tokai, Ibaraki 319-1195, Japan}

\author{Yuichi Ohnuma}
\affiliation{Research Center for Advanced Science and Technology (RCAST),
The University of Tokyo, Meguro, Tokyo 153-8904, Japan
}

\author{Se Kwon Kim}
\affiliation{
Department of Physics, Korea Advanced Institute of Science and Technology, Daejeon 34141, Republic of Korea}

\date{\today}

\begin{abstract}
The celebrated Wiedemann-Franz (WF) law which governs the relation between charge and heat transport traces back to the experimental discovery in 1853 by Wiedemann and Franz. Despite the fundamental difference of the quantum-statistical properties between fermions and bosons, the linear-in-$T$ behavior of the WF law at low temperatures has recently been found to be the universal property by the discovery of the WF law for magnon transport. However, the WF law is for the linear response, and whether or not the universal law is valid even in the nonlinear regime of Bose systems remains an open issue. Here we provide a solution to this fundamental challenge. We show that the ratio of the thermal to spin transport coefficient of magnons in topologically trivial insulating magnets exhibits a different behavior from the linear response and the universal law breaks down in the strong nonlinear regime. This finding is within experimental reach with current device and measurement technologies. Our discovery is the key ingredient in magnon-based spintronics, in the evaluation of the figure of merit for thermomagnetic conversion elements of spintronics devices.
\end{abstract}

\maketitle

\section{Introduction}
\label{sec:Intro}

The research on thermoelectric properties of materials 
started more than two centuries ago and it has a long history.
The celebrated Wiedemann-Franz (WF) law
which dictates the linear relation between charge and heat transport
traces back to the experimental discovery in 1853 
by Wiedemann and Franz~\cite{WFgermany}
that the ratio of the thermal $\kappa$
to electrical conductivity $\sigma$
of several metals reduces to approximately the same value 
for a fixed temperature.
Lorenz established that this ratio is linear in the absolute temperature $T$
and 
the proportionality constant 
takes a material-independent value~\cite{Lorenz1881}.
Using quantum theory on solids, 
Sommerfeld appropriately derived the universal constant, 
dubbed the Lorenz number ${\cal{L}}_{\text{e}}$,
which is independent of any material parameters~\cite{Sommerfeld1928}.
Thus the WF law has been formulated~\cite{LandauWF,AMermin}
{at low temperature}
in the form of
\begin{align}
 \frac{\kappa}{\sigma} &\stackrel{\rightarrow }{=} 
 {\cal{L}}_{\text{e}} T,
    \nonumber \\
{\cal{L}}_{\text{e}}&:= \frac{{\pi}^2}{3} \Big(\frac{{k_{\text{B}}}}{e}\Big)^2,
   \nonumber 
\end{align}
where
$e$ is the elementary charge
and $ k_{\text{B}} $ the Boltzmann constant.
This universal law characterizes the figure of merit for thermoelectric conversion elements and has been playing a central role in electronics.

Toward efficient transmission of information 
that goes beyond what is offered by conventional electronics,
the last two decades have seen a rapid development of spintronics,
aiming at utilizing another degrees of freedom of electrons, spins,
by means of spin transport~\cite{MagnonSpintronics,ReviewMagnon}.
For this holy grail,
it is desirable to formulate the spin analog of the WF law
because the law is expected to be a promising building block
in spintronics,
in the evaluation of the figure of merit 
for thermomagnetic conversion elements of spintronics devices~\footnote{
{We refer to the spin analog  of the figure of merit 
for thermoelectric conversion elements
as that for thermomagnetic ones $Z_{\text{s}}$,
i.e.,
$Z_{\text{s}}:= {\mathcal{S}}^2 G/K$,
where
${\mathcal{S}} $, $G$, and $K$ are
the spin Seebeck coefficient,
the spin conductivity,
and the thermal conductivity of magnons, respectively.}}.
Then we have established 
the magnon analog of the WF law, 
namely,
the WF law for magnon transport, 
in ferromagnets and antiferromagnets~\cite{magnonWF,KSJD,KJD}. 
Magnons are bosonic magnetic excitations, i.e., the quantized spin-waves.
Since magnons carry the spin angular momentum,
spin currents are generated in the absence of charge currents
when magnons propagate in insulating magnets.
Thus the WF law for magnon transport, dubbed the magnonic WF law, 
is defined as the ratio of the thermal conductivity $K_1$ 
within the linear response regime
to spin conductivity $G$ of magnons.
In the bulk of topologically trivial insulating magnets 
{at sufficiently low temperatures compared to the magnon energy gap,}
the ratio reduces to~\cite{KSJD}
\begin{subequations}
\begin{align}
 \frac{K_1}{G} &\stackrel{\rightarrow }{=} 
 {\cal{L}}_{1} T,
   \label{eqn:WF1}  \\
   {\cal{L}}_{1}&:= \frac{5}{2} \Big(\frac{{k_{\text{B}}}}{g \mu _{\text{B}}}\Big)^2,
   \label{eqn:Lorenz}
\end{align}
\end{subequations}
where
$g$ is the $g$-factor of the constituent spins
and $\mu _{\text{B}}$ the Bohr magneton.
The thermomagnetic ratio is linear in temperature.
In analogy to charge transport in metals,
we refer to this behavior 
as the magnonic WF law.
The constant $ {\cal{L}}_{1}$ 
analogous to the Lorenz number,
i.e., the magnetic Lorenz number of magnons,
is independent of any material parameters
except the $g$-factor,
which is material specific.
The role of the charge $e$ is played by $g \mu_{\text{B}}$.

Magnons are bosonic excitations,
while electrons are fermions.
Still, remarkably, 
the magnonic WF law exhibits
the same linear-in-$T$ behavior at low temperatures 
as the one for electronic transport 
despite the fact that 
the quantum-statistical properties of bosons and fermions 
are fundamentally different, 
particularly in the low temperature regime where quantum effects dominate.
In that sense, the linear-in-$T$ behavior of the WF law 
is found to be the universal property.

As spintronics technologies develop, 
there has recently been a growing interest 
in the properties of the nonlinear response~\cite{NonlinearSpinSeebeck,NonlinearAndo2014,Nonlinear2020,NonlinearSpinCurrent,KatsuraNonlinearSpinDrude,NakaiNagaosa2019,NonlinearNernst2019,NonlinearWF2020,OshikawaNonlinear2020}.
However, 
the existing theory for the magnonic WF law is intended 
to be applied only for the linear response regime.
Whether or not the universal law is valid 
even in the nonlinear regime remains an open issue.
In this paper, 
we provide a solution to this fundamental challenge 
by using the Boltzmann equation.
This is the main aim of this paper.

This paper is organized as follows.
In Sec.~\ref{sec:2}
we investigate longitudinal thermal transport of magnons
in the bulk of insulating magnets, 
{and study the validity and violation of the magnonic WF law 
in the nonlinear response regime.}
Then, we remark on several issues in Sec.~\ref{sec:Discussion}.
Finally,
we give an estimate for the experimental feasibility
in Sec.~\ref{sec:Estimate}
and 
summarize in Sec.~\ref{sec:summary}.
Technical details are described in the Appendix.

\section{Nonlinear thermal transport}
\label{sec:2}

We consider longitudinal transport of magnons
in the bulk of a topologically trivial
three-dimensional insulating magnet~\footnote{
We refer to the magnet where Berry curvatures are zero
as the topologically trivial magnet.}, 
subjected to a temperature gradient,
where
the magnon of the energy dispersion relation 
$ \epsilon _{{{\mathbf{k}} }} =Dk^2+\Delta  $
with the group velocity
$ {\mathbf{v}}_{{\mathbf{k}}} = 
\partial \epsilon _{{\mathbf{k}} }/(\partial \hbar {\mathbf{k}})$
carries the spin angular momentum $-1$
in units of 
the reduced Planck constant
$\hbar$:
In which
$k:=|{\mathbf{k}}|$
denotes the magnitude of the wavenumber
$ {\mathbf{k}}=(k_x, k_y, k_z) $,
$D$ represents the spin stiffness constant,
and $\Delta$ 
is the magnon energy gap,
e.g., due to an external magnetic field, a spin anisotropy, etc.
Throughout this paper
assuming 
that the magnon energy gap takes a nonzero value $ \Delta \neq 0 $
and
that the nonequilibrium Bose distribution function of magnons 
$ f_{{\mathbf{k}} }$
is described by the Boltzmann equation 
within the quasiparticle approximation,
we study magnon transport at low temperatures
using a relaxation time approximation~\cite{seeAppendices}.
Note that if one assumed a magnon energy dispersion
including the $k$-linear term,
nonreciprocal responses~\cite{ReviewNonreciprocal},
e.g., the Doppler shift of spin-waves~\cite{DopplerScience},
could arise in certain materials
with broken inversion symmetry,
which is outside the scope of this paper:
See Ref.~\cite{SKKNonreciprocitySpinWaves}
for the generation of the magnon nonreciprocity
in the absence of a finite energy gap, 
i.e., nonreciprocal transport of gapless spin-waves.

The applied temperature gradient,
$ {\mathbf{\nabla}} T=({\text{const.}})  $,
induces a magnonic spin current along the longitudinal direction,
which leads to an accumulation of magnons at the boundaries
and builds up a nonuniform magnetization in the magnet.
This magnetization gradient plays a role of 
an effective magnetic field gradient
$ {\mathbf{\nabla}} B $,
which works as the gradient of a nonequilibrium spin chemical potential~\cite{Basso2,MagnonChemicalWees,YacobyChemical,demokritov},
and drives magnon currents~\cite{Haldane2,magnon2,Fujimoto,KSJD}.
Therefore, magnon transport 
subjected to the temperature gradient
along the $x$ axis, 
$\partial _x T=({\text{const.}})$,
is characterized
including the nonlinear response
by the longitudinal transport coefficient
$L_{ij}$ 
for
$i \in {\mathbb{N}}$
and
$j \in {\mathbb{N}}$
as
\begin{eqnarray}
\begin{pmatrix}
 j_{x }^{\text{s}}   \\ 
 j_{x }^{\text{E}} 
\end{pmatrix}
=
\begin{pmatrix}
L_{11} & L_{12} & L_{13} & L_{14} & L_{15} & L_{16} \\ 
L_{21} & L_{22} & L_{23} & L_{24} & L_{25} & L_{26}
\end{pmatrix}
\begin{pmatrix}
  \partial _x B  \\   
  - \frac{\partial _x T}{T} \\
  (\partial _x B)^2 \\
   - (\frac{\partial _x T}{T}) (\partial _x B)  \\
   (\frac{\partial _x T}{T})^2  \\
     -(\frac{\partial _x T}{T})^3   \nonumber
\end{pmatrix},
\label{eqn:matrix}
\\
\end{eqnarray}
where 
the spin current density
$ {\mathbf{j}}^{\text{s}} 
=(j_{x }^{\text{s}}, j_{y }^{\text{s}}, j_{z }^{\text{s}} )  $
and the energy current density
${\mathbf{j}}^{\text{E}}
=(j_{x }^{\text{E}}, j_{y }^{\text{E}}, j_{z }^{\text{E}})$
are defined as~\cite{KSJD}
\begin{subequations}
\begin{align}
{\mathbf{j}}^{\text{s}} :=& -\int  \frac{d^3{\mathbf{k}}}{(2 \pi)^3}
g \mu_{\text{B}} {\mathbf{v}}_{{\mathbf{k}}} f_{{\mathbf{k}} } ,
  \label{eqn:js}  \\
   {\mathbf{j}}^{\text{E}} :=& \int  \frac{d^3{\mathbf{k}}}{(2 \pi)^3}
   \epsilon _{{\mathbf{k}} }   {\mathbf{v}}_{{\mathbf{k}}} f_{{\mathbf{k}}} ,
  \label{eqn:jE}
\end{align}
\end{subequations}
respectively.
The transport coefficient $L_{11}$
is identified with the spin conductivity of magnons~\cite{magnon2}
$G:=L_{11}$.
In contrast to the junction system~\cite{magnonWF,KNmagnonNoiseJunction},
the second-order response vanishes~\footnote{
This result changes in general 
if one assumes a magnon energy dispersion
with the $k$-linear term.}
in the bulk of topologically trivial insulating magnets
due to the property of the odd function in $k_x$~\cite{seeAppendices}:
\begin{align}
 L_{13}= L_{14}= L_{15}= L_{23}=  L_{24}= L_{25}= 0.
  \label{eqn:secondorder}
\end{align}
{Note that 
since the third-order response to the temperature gradient
$ O\big((\partial _x T)^3\big) $
can become dominant at sufficiently low temperatures 
compared to the magnon gap~\cite{seeAppendices},} 
we neglect the other third-order terms
such as 
$ O\big((\partial _x T)^2(\partial _x B)^1\big) $,
$ O\big((\partial _x T)^1(\partial _x B)^2\big) $,
and
$ O\big((\partial _x B)^3\big) $.

In analogy to charge transport~\cite{AMermin}
and
using Eq.~\eqref{eqn:matrix},
we formulate thermal transport of magnons
in the nonlinear response regime.
Under the applied temperature gradient,
the magnonic spin current is generated and
this leads to an accumulation of magnons at the boundaries.
Consequently, the nonuniform magnetization is developed
and this effective magnetic field gradient
$  \partial _x B $
generates a counter-current of magnons.
Then, the system reaches a stationary state such that 
in- and out-flowing magnonic spin currents balance each other
$ j_{x }^{\text{s}} =0$,
which results in
$  \partial _x  B= \partial _x  B^{\ast }$
with
\begin{align}
   \partial _x  B^{\ast } :=  \frac{L_{12 }}{L_{11 }} 
   \frac{ \partial _x  T}{T}
   + \frac{L_{16}}{L_{11 }} 
   \Big(\frac{ \partial _x  T}{T}\Big)^3.
  \label{eqn:Beff}
\end{align}
Thus the thermal conductivity is measured.
This effective magnetic field gradient in the new quasiequilibrium state 
$ \partial _x  B^{\ast }  $
brings the nonequilibrium spin chemical potential~\cite{Basso2},
being peculiar to the system out of equilibrium,
and contributes to the thermal conductivity 
associated with the heat current density~\cite{KSJD,Basso2}, 
$j_x^Q
=L_{21}  \partial _xB^{\ast}
-L_{22} {\partial _xT}/{T}
-L_{26} ({\partial _xT}/{T})^3
$,
as
\begin{align}
j_x^Q= - K_1  \partial _xT
- K_2 (\partial _xT)^2
- K_3 (\partial _xT)^3,
  \label{eqn:JQ}
\end{align}
where
\begin{subequations}
\begin{align}
K_1&:=\frac{1}{T}\Big(L_{22}-\frac{L_{12 }L_{21}}{L_{11 }} \Big),
\label{eqn:K1}
\\
K_2&=0,
\label{eqn:K2}
\\
K_3&:=\frac{1}{T^3}\Big(L_{26}-\frac{L_{16 }L_{21}}{L_{11 }} \Big),
\label{eqn:K3}
\end{align}
\end{subequations}
and
$K_1$
represents the thermal conductivity in the linear response regime
and 
$K_{2(3)}$ 
is
the thermal transport coefficient of the 
second-order (third-order) nonlinear response.
Note that in the stationary state under the applied temperature gradient,
the heat current density
$  j_x^Q $
is different from the energy current density
$ j_{x }^{\text{E}} $~\cite{KSJD,Basso2,QBEmagnon},
\begin{align}
 j_x^Q  \neq  j_{x }^{\text{E}} ,
 \label{eqn:JQJE}
\end{align}
in that
\begin{subequations}
\begin{align}
j_x^Q=&
L_{21}  \partial _xB^{\ast}
-L_{22} \frac{\partial _xT}{T}
-L_{26} \Big(\frac{\partial _xT}{T}\Big)^3, 
\label{eqn:JQ2}
\\
j_{x }^{\text{E}} =&
\   \   \   \   \   
0
\   \   \   \   \  
-L_{22} \frac{\partial _xT}{T}
-L_{26} \Big(\frac{\partial _xT}{T}\Big)^3.
\label{eqn:JE2}
\end{align}
\end{subequations}

We remark that if one wrongly omits the contribution of 
$ \partial _xB^{\ast}$
associated with the counter-current
and identifies 
$  j_{x }^{\text{E}}$
as the heat current density
in theoretical calculation,
the ratio of the thermal to spin conductivity
would not obey the magnonic WF law,
breaking the linear-in-$T$ behavior,
even in the linear response regime~\cite{KSJD}.
Note that for thermal transport of electrons in metals,
the contribution of the counter-current
is strongly suppressed by the sharp Fermi surface of fermions
at temperatures $ k_{\text{B}} T$,
which is much smaller than the Fermi energy
even at room temperature.
This is the crucial difference in the thermal conductivity
between magnons and electrons,
i.e.,
bosons and fermions, respectively.

We evaluate the thermal transport coefficient of 
the third-order nonlinear response $K_3 $.
The Onsager relation holds   
$L_{12}=L_{21}$
and
at low temperatures $k_{\text{B}}T  \ll  \Delta $,
it reduces to~\cite{KSJD}
$  L_{12}/L_{11}=L_{21}/L_{11}
\stackrel{\rightarrow }{=}  -{\Delta}/({g \mu _{\text{B}}}) $.
Thus, the thermal transport coefficient 
of the third-order nonlinear response
at low temperature is recast into
\begin{align}
 K_3 \stackrel{\rightarrow }{=} 
\frac{1}{T^3}
\Big(L_{26}+\frac{{\Delta}}{{g \mu _{\text{B}}}}L_{16}
\Big).
\label{eqn:K3II}
\end{align}
A straightforward calculation using the Boltzmann equation
provides
the transport coefficients of the nonlinear response
at low temperatures
as~\cite{seeAppendices}
\begin{subequations}
\begin{align}
 L_{16}& \stackrel{\rightarrow }{=} 
 -g \mu _{\text{B}}F {\text{e}}^{-b}
\Big[
\frac{{\Delta}^3}{(\beta D)^{7/2}}A_3
+\frac{3D{\Delta}^2}{(\beta D)^{9/2}}A_4
+O(T^{11/2})
\Big],  
\label{eqn:L16}
\\
 L_{26}& \stackrel{\rightarrow }{=} 
F {\text{e}}^{-b}
\Big[
\frac{{\Delta}^4}{(\beta D)^{7/2}}A_3
+\frac{4D{\Delta}^3}{(\beta D)^{9/2}}A_4 
+O(T^{11/2})
\Big],  
\label{eqn:L26}
\end{align}
\end{subequations}
where
$A_n:={\sqrt{\pi}}(2n)!/[2^{2n+1}(n!)] $
is the Gaussian integral for $n \in {\mathbb{N}}$,
$F:=({2D}/{\hbar})^4[{{\tau}^3{\beta}^3}/({{10}{\pi}^2})]$,
the inverse temperature $\beta :=1/(k_{\text{B}}T)$,
$b:= \beta \Delta $,
and
the relaxation time $\tau$.
Note that at low temperatures
the relaxation time
takes a constant value of being temperature-independent:
At sufficiently low temperatures,
the effect of magnon-magnon interactions and
that of phonons 
are negligibly small,
and
impurity scattering 
makes a major contribution to the relaxation.
Under the assumption that 
impurities are dilute
and 
scattering is elastic and spatially isotopic
with the impurity potential localized in space,
the relaxation time 
at low temperatures
reduces to~\cite{QBEmagnon}
$ \tau \stackrel{\rightarrow }{=} {\hbar}/({2\alpha\Delta})  $,
where
$\alpha$ is the Gilbert damping constant.
Since the Gilbert damping constant is little influenced by temperature~\cite{LLGspintroReview}
(i.e., the temperature dependence is negligibly small),
the relaxation time at low temperatures
takes a constant value of being temperature-independent.

At low temperatures
the thermal transport coefficient of the third-order nonlinear response
reduces to 
\begin{align}
 K_3 \stackrel{\rightarrow }{=}
\frac{1}{T^3} F {\text{e}}^{-b}
\frac{D{\Delta}^3}{(\beta D)^{9/2}}A_4,
\label{eqn:K3III}
\end{align}
and
the spin conductivity to
$ G \stackrel{\rightarrow }{=}
(g \mu _{\text{B}})^2 {\text{e}}^{-b} \tau
{(k_{\text{B}}T)^{3/2}}/({4{\pi}^{3/2}{\hbar}^2\sqrt{D}}) $.
Thus, we find at low temperatures $k_{\text{B}}T  \ll  \Delta $
that the ratio of the thermal transport coefficient 
of the third-order nonlinear response
to the spin conductivity
is given as
\begin{subequations}
\begin{align}
 \frac{K_3}{G} &\stackrel{\rightarrow }{=} 
 {\cal{L}}_{3} \frac{1}{T^3},
   \label{eqn:WF3}   \\
{\cal{L}}_{3} &:=
\frac{32 {\tau}^2 D {\Delta}^3 A_4}{{5}{\sqrt{\pi}}{\hbar}^2
 (g \mu _{\text{B}})^2}.
 \label{eqn:L3}
\end{align}
\end{subequations}
The thermomagnetic ratio in the nonlinear regime
is proportional to $1/T^3$ [Eq.~\eqref{eqn:WF3}].
This is in contrast to the one
in the linear response regime
$K_1/G$,
which exhibits the linear-in-$T$ behavior [Eq.~\eqref{eqn:WF1}].
The proportionality constant ${\cal{L}}_{3}  $,
which is independent of temperature,
is less universal than 
the magnetic Lorenz number
${\cal{L}}_{1}  $
[Eq.~\eqref{eqn:Lorenz}]
in that the constant depends on other material parameters 
as well as the $g$-factor,
such as 
the spin exchange interaction,
the spin anisotropy,
the spin length,
the lattice constant, etc.
Note that each transport coefficient
does not diverge even at low temperature $ T \stackrel{\rightarrow }{=} 0 $
as
$ K_3 \propto  {\text{e}}^{-b}/T^{3/2}  \stackrel{\rightarrow }{=} 0  $
and 
$ G \propto  T^{3/2} {\text{e}}^{-b} 
\stackrel{\rightarrow }{=} 0  $.


Finally, we discuss the validity and violation
of the magnonic WF law 
in the nonlinear response regime.
The magnonic WF law is originally for the linear response 
[Eq.~\eqref{eqn:WF1}].
Instead of the thermal conductivity $K_1$,
it is recast in terms of the heat current density 
$j_x^Q= - K_1  \partial _xT$
as
$  {j_x^Q}/{G} = - [{\cal{L}}_{1} ( \partial _xT)]T   $,
which states that the ratio of the heat current density 
to the magnonic spin conductivity $G$
is linear in temperature 
for a fixed temperature gradient
$( \partial _xT)=(\text{const.})$.
This is the magnonic WF law in terms of the heat current density.
Since the heat current density includes the nonlinear response
[Eq.~\eqref{eqn:JQ}],
it can be concluded that the magnonic WF law does hold
even in the nonlinear response regime
if the ratio of the heat current density
to the magnonic spin conductivity $G$
exhibits the linear-in-$T$ behavior
for the fixed temperature gradient: This is the criterion 
for the magnonic WF law in the nonlinear response regime.

From Eq.~\eqref{eqn:JQ}
the ratio of the heat current density
including the nonlinear response
to the magnonic spin conductivity $G$ becomes
\begin{align}
\frac{j_x^Q}{G}
= - \frac{K_1}{G} ( \partial _xT)
- \frac{K_2}{G} ( \partial _xT)^2
- \frac{K_3}{G} ( \partial _xT)^3.
\label{eqn:JQGII}
\end{align}
Since the thermal transport coefficient 
of the second-order nonlinear response vanishes 
$K_2=0$ [Eq.~\eqref{eqn:K2}],
the magnonic WF law holds even in the nonlinear regime if the ratio of the thermal transport coefficient of the third-order nonlinear response
to the spin conductivity, $K_3/G$, 
exhibits the linear-in-$T$ behavior.
However, we find from Eq.~\eqref{eqn:WF3}
that the ratio $K_3/G$ is proportional to $1/T^3$ and does not exhibit the linear-in-$T$ behavior.
Thus, it is concluded that 
the magnonic WF law violates in the nonlinear regime.

We remark that
the magnonic WF law breaks down
in the strong nonlinear regime
where the third-order nonlinear response 
to the temperature gradient
$O\big(( \partial _xT)^3\big)$
becomes relevant
by the large temperature gradient.
It is not until 
the third-order response contributes
that the law violates.
Since the thermal transport coefficient of 
the second-order nonlinear response vanishes $K_2=0$,
the universal law remains valid even in the nonlinear regime
where the temperature gradient is large but not enough
for the third-order response to become relevant:
We refer to this region as the weak nonlinear regime
for convenience.
In conclusion,
in the bulk of topologically trivial insulating magnets,
the magnonic WF law remains valid even in the weak nonlinear regime
but breaks down in the strong nonlinear regime.

\section{Discussion}
\label{sec:Discussion}

In contrast to the bulk of 
topologically trivial materials studied in this paper
[Eq.~\eqref{eqn:secondorder}],
the second-order nonlinear response does not vanish
in junction systems~\cite{magnonWF,KNmagnonNoiseJunction}, 
including quantum dot systems,
and it contributes to thermal transport.
Therefore,
in the quantum dot system~\cite{NLelWFviolation,ViolationWF_QD}
the WF law for electronic transport violates
in the weak nonlinear regime
due to the second-order response.
Thus, we find that 
in the bulk of topologically trivial materials
the WF law is more robust against the nonlinear effect
compared with in junction systems,
in that 
the law breaks down in the strong nonlinear regime 
for the bulk of topologically trivial materials,
while it violates in the weak nonlinear regime for junction systems.

Note that throughout this paper,
we focus on longitudinal thermal transport in the bulk of 
topologically trivial insulating magnets
where Berry curvatures are zero,
and find that the WF law breaks down
in the strong nonlinear regime.
In the bulk of topological materials~\footnote{See Ref.~\cite{kondoNonlinear} for topological magnon systems.}, however,
the WF law for electronic Hall transport violates
in the weak nonlinear regime 
due to the second-order response
arising from nonzero Berry curvatures~\cite{NonlinearWF2020}.
Thus, it is concluded that 
in the bulk of topologically trivial materials
the WF law is more robust against the nonlinear effect
compared with in the bulk of topological materials.

In this paper,
we have studied magnon transport 
in the nonlinear response regime
under the assumption that 
the energy dispersion of magnons
is gapped and parabolic
in terms of $k$.
We remark that 
the second-order nonlinear response [Eq.~\eqref{eqn:secondorder}]
does not vanish in general
if one assumes a magnon energy dispersion
including the $k$-linear term.
In that case,
the magnon nonreciprocity~\cite{ReviewNonreciprocal},
e.g., the spin-wave Doppler shift~\cite{DopplerScience},
could arise in certain materials
with broken inversion symmetry.
See Ref.~\cite{SKKNonreciprocitySpinWaves}
for nonreciprocal transport in a gapless spin-wave system,
i.e., the magnon nonreciprocity
in the absence of a finite energy gap.

In general, there exists a spin anisotropy in magnets,
which causes the magnon energy gap.
Therefore, 
toward the development of various functions of spintronics devices,
it is of importance to establish the fundamental principle 
of magnon transport even in the gapped systems.
Hence, we studied the gapped magnonic systems.
In the gapped system, 
the magnonic WF law holds only at low temperature
$k_{\text{B}}T  \ll  \Delta $ 
and the linear-in-$T$ behavior violates at higher temperatures 
even within the linear response~\cite{magnonWF,KSJD}.
Therefore, in this paper focusing on the gapped magnonic system
at such low temperatures,
we have studied the effect of the nonlinear response
on the magnonic WF law 
(i.e., the linear-in-$T$ behavior).
We also remark that 
if the magnon gap is much smaller than the thermal energy, 
then there would be a large number of low-energy magnons 
and therefore frequent interactions between magnons. 
In this case, we would need to consider the magnon-magnon interaction 
to capture the transport properties well. 
In this work, we focused on the systems with the magnon gap 
larger than the thermal energy in part so that 
the magnon-magnon interactions can be neglected. 
Studying the effect of magnon-magnon interactions is 
beyond the scope of our current work.
For these reasons, in this paper, 
we have focused on the gapped magnonic system 
at low temperatures.
Still, 
it will be of significance to develop our work 
into the gapless magnon mode that possesses appropriate symmetry
and include the magnon-magnon interaction.
We leave the advanced study for future work.

By manipulating the applied magnetic field, 
magnon and phonon thermal conductivities 
can be distinguished experimentally~\cite{PhononMagnonPrasai} 
because the former does depend strongly on the field 
whereas the latter does not. 
In fact, in Ref.~\cite{PhononMagnonPrasai}, 
it has been experimentally shown that 
thermal conductivities of phonons and magnons in a magnetic insulator 
can be separately characterized at low temperatures $T \leq O(1)$K. 
For this reason, we focus only on magnon thermal conductivity in our work.

\section{Estimates for experiments}
\label{sec:Estimate}

In the bulk of topologically trivial insulating magnets,
the heat current density [Eq.~\eqref{eqn:JQ}] 
consists of the linear response 
$ K_1 \partial _xT  $
and the third-order nonlinear response
$ K_3 ( \partial _xT)^3  $.
When 
the large temperature gradient is applied 
enough that 
the third-order nonlinear response 
begins to contribute
in that 
the ratio of the nonlinear to linear response amounts to $1/10$,
$ K_3 ( \partial _xT)^3 / (K_1 \partial _xT)=1/10$,
we identify it with the strong nonlinear regime.
Thus, for observation of the violation of the magnonic WF law
in the bulk of insulating magnets,
the temperature gradient needs to reach
$ | \partial _xT | =\sqrt{(1/10)(K_1/K_3)}=:\mathcal{T}$,
where
$ K_1/K_3={25}{\sqrt{\pi}}{\hbar}^2 k_{\text{B}}^2T^4/(64 {\tau}^2 D {\Delta}^3 A_4) $.
The criterion for observation of the violation is
whether or not
the temperature gradient exceeds 
the value of $\mathcal{T}$.

For an estimate, 
we assume the following experimental parameter values~\cite{CrOexp,CrOexp2} 
for ${\text{Cr}}_2{\text{O}}_3$:
$D= 10 $ meV(nm)$^2$,
$ {\Delta} \sim 4$ meV,
$ \alpha= O(10^{-3})$,
and
$T=4$ K.
This results in
$ {\mathcal{T}}=O(10) $ K/mm.
In addition, 
since the value of ${\mathcal{T}}$ is proportional 
to the Gilbert damping constant
and it takes~\cite{LLGspintroReview,GilbertInsulator}
$ \alpha= O(10^{-4})$ or even $ \alpha < O(10^{-4})$
for YIG depending on the shape,
it is roughly estimated as 
$  {\mathcal{T}} = O(1) $ K/mm 
or even
$  {\mathcal{T}} < O(1) $ K/mm for YIG,
respectively.
In both cases, 
those temperature gradients are
experimentally realizable~\cite{ChudoPrivate}.

Observation of 
long-distance transport of spin-wave spin currents~\cite{WeesNatPhys}
and 
measurement of
the nonequilibrium spin chemical potential~\cite{YacobyChemical},
the magnonic spin conductivity~\cite{MagnonG,MagnonChemicalWees},
and the thermal conductivity~\cite{onose}
have been reported.
Refs.~\cite{tabuchi,tabuchiScience,magnon10mK}
develop magnonics technologies at low temperatures.
Given these estimates,
we expect that observation of 
the magnonic WF law and the violation, 
while being challenging, 
seem within experimental reach 
with current device and measurement techniques.
{The key is to decrease the overall temperature 
while the temperature gradient is maintained constant.}

\section{Summary}
\label{sec:summary}

Focusing on longitudinal transport of magnons at low temperatures 
in the bulk of topologically trivial insulating magnets,
we have studied the validity and have found the violation of
the magnonic Wiedemann-Franz law
in the nonlinear response regime.
In terms of the heat current density,
the magnonic Wiedemann-Franz law is recast into the form that 
the ratio of the heat current density 
to the magnonic spin conductivity is linear in temperature 
for a fixed temperature gradient.
Then we have shown that the universality of the Wiedemann-Franz law, 
the linear-in-$T$ behavior, breaks down 
in the strong nonlinear regime.
In contrast to the linear response,
the ratio of the thermal transport coefficient of the third-order 
nonlinear response to the spin conductivity is proportional to $1/T^3$,
and the proportionality constant is less universal in that 
it depends on other material parameters as well as the $g$-factor.
Thus, the universal law violates in the strong nonlinear regime 
where the third-order nonlinear response becomes relevant
by the large temperature gradient.
Since the second-order nonlinear response vanishes 
in the bulk of topologically trivial insulating magnets,
the magnonic Wiedemann-Franz law 
remains valid even in the weak nonlinear regime
but
breaks down in the strong nonlinear regime.
Those findings are within experimental reach with current device 
and measurement technologies.
Toward efficient transmission of information
that goes beyond what is offered by conventional electronics,
our discovery is a promising building block
in magnon-based spintronics,
in the evaluation of the figure of merit 
for thermomagnetic conversion elements of spintronics devices.

\acknowledgments

The author (K. N.) would like to thank D. Loss
for the collaborative work on the related study
through fruitful discussions.
We are grateful also to 
Y. Araki for useful discussions
and
H. Chudo for helpful feedback on the experimental feasibility.
We acknowledge support
by JSPS KAKENHI Grant Number JP20K14420 (K. N.)
and JP22K03519 (K. N.),
by Leading Initiative for Excellent Young Researchers, MEXT, Japan (K. N.),
and 
by JST ERATO Grant No. JPMJER1601 (Y. O.).
The author (S. K. K.) is supported by Brain Pool Plus Program through the National Research Foundation of Korea funded by the Ministry of Science and ICT (Grant No. NRF-2020H1D3A2A03099291) and by the National Research Foundation of Korea funded by the Korea Government via the SRC Center for Quantum Coherence in Condensed Matter (Grant No. NRF-2016R1A5A1008184).

\begin{widetext}

\appendix*

\section{Transport coefficients of magnons 
in the nonlinear response regime}
\label{sec:AppendixA}

In this Appendix,
we provide some details of the straightforward calculation
for the transport coefficient $L_{ij}$
of the nonlinear response
in the bulk of topologically trivial insulating magnets,
and make a few remarks.
Under the relaxation time approximation,
the Boltzmann equation of the quasiparticle approximation
describes the transport property 
of a steady state in terms of time as~\cite{mahan,haug,Haldane2,magnon2,Fujimoto,Basso2}
\begin{align}
\Big({\mathbf{v}}_{{\mathbf{k}}} \cdot {\mathbf{\nabla }}T
\frac{\partial}{\partial T}
- g \mu_{\text{B}} {\mathbf{v}}_{{\mathbf{k}}}\cdot {\mathbf{\nabla }}B 
\frac{\partial}{\partial  \epsilon _{{{\mathbf{k}} }}}
\Big)
f_{{\mathbf{k}} }
=-\frac{f_{{\mathbf{k}} } - f^0_{{\mathbf{k}} }}{{\tau}}, 
\end{align}
where $\tau$ is the relaxation time,
the Bose distribution function of magnons out of equilibrium
$f_{{\mathbf{k}} }$,
and the one in equilibrium
$ f^0_{{\mathbf{k}} } := ({\text{e}}^{\beta  \epsilon _{{{\mathbf{k}} }}}-1)^{-1}   $.
Defining the deviation from equilibrium
$ g_{{\mathbf{k}} } :=  f_{{\mathbf{k}} } - f^0_{{\mathbf{k}} }  $,
it is described as 
\begin{align}
 g_{{\mathbf{k}} }= -\tau
\Big({\mathbf{v}}_{{\mathbf{k}}} \cdot {\mathbf{\nabla }}T
\frac{\partial}{\partial T}
- g \mu_{\text{B}} {\mathbf{v}}_{{\mathbf{k}}}\cdot {\mathbf{\nabla }}B 
\frac{\partial}{\partial  \epsilon _{{{\mathbf{k}} }}}
\Big)
(f^0_{{\mathbf{k}} } + g_{{\mathbf{k}} }).
\end{align}
Using the method of successive substitution for 
$  g_{{\mathbf{k}} } $,
the deviation up to $O({\tau}^3)$ 
is given as
\begin{subequations}
\begin{align}
g_{{\mathbf{k}} }=&g_1+g_2+g_3+O({\tau}^4),  \\
g_1:=&
-\tau\Big({\mathbf{v}}_{{\mathbf{k}}} \cdot {\mathbf{\nabla }}T
\frac{\partial}{\partial T}
- g \mu_{\text{B}} {\mathbf{v}}_{{\mathbf{k}}}\cdot {\mathbf{\nabla }}B 
\frac{\partial}{\partial  \epsilon _{{{\mathbf{k}} }}}
\Big)
f_{{\mathbf{k}} }^0,  \\
g_2:=&
{\tau}^2 \Big({\mathbf{v}}_{{\mathbf{k}}} \cdot {\mathbf{\nabla }}T
\frac{\partial}{\partial T}
- g \mu_{\text{B}} {\mathbf{v}}_{{\mathbf{k}}}\cdot {\mathbf{\nabla }}B 
\frac{\partial}{\partial  \epsilon _{{{\mathbf{k}} }}}
\Big)^2
f_{{\mathbf{k}} }^0,  \\
g_3:=&
-{\tau}^3 \Big({\mathbf{v}}_{{\mathbf{k}}} \cdot {\mathbf{\nabla }}T
\frac{\partial}{\partial T}
- g \mu_{\text{B}} {\mathbf{v}}_{{\mathbf{k}}}\cdot {\mathbf{\nabla }}B 
\frac{\partial}{\partial  \epsilon _{{{\mathbf{k}} }}}
\Big)^3
f_{{\mathbf{k}} }^0,
\end{align}
\end{subequations}
where
$ g_1= O({\tau}) $,
$ g_2= O({\tau}^2) $,
and
$ g_3= O({\tau}^3) $.
From
$ \partial/(\partial T)=- (\epsilon _{{{\mathbf{k}} }}/T) 
[\partial/(\partial \epsilon _{{{\mathbf{k}} }})]$,
the component $g_3$ is recast into
\begin{align}
g_3=
{\tau}^3 \Big({\mathbf{v}}_{{\mathbf{k}}} \cdot {\mathbf{\nabla }}T
\frac{\epsilon _{{{\mathbf{k}} }}}{T}
+ g \mu_{\text{B}} {\mathbf{v}}_{{\mathbf{k}}}\cdot {\mathbf{\nabla }}B 
\Big)^3
\frac{{\partial}^3 f_{{\mathbf{k}} }^0}{\partial  {\epsilon _{{{\mathbf{k}} }}}^3}.
\end{align}
{Focusing on sufficiently low temperatures 
compared to the magnon gap
and thus assuming 
$  \Delta/(k_{\text{B}}T)  \gg   
g \mu_{\text{B}} \partial_x B/(k_{\text{B}} \partial_x T) $,
the third-order response to the temperature gradient
$ O\big(({\mathbf{\nabla }}T)^3\big) $
becomes dominant.}
{This condition can be met 
when the system has a sufficiently large magnetic anisotropy 
and the resulting magnon gap,
such as
$\Delta \sim 4$ meV $=O(10)$ T
for~\cite{CrOexp,CrOexp2,KSJD} 
${\text{Cr}}_2{\text{O}}_3$,
$\Delta \sim 2$ meV $=O(10)$ T
for~\cite{SrRuO3} 
${\text{SrRu}}{\text{O}}_3$,
and
$\Delta \sim 0.4$ meV $=O(1)$ T
for~\cite{CrI3} 
${\text{Cr}}{\text{I}}_3$.}
Then, the component reduces to
\begin{align}
g_3 \stackrel{\rightarrow }{=}
-{\tau}^3 ({\mathbf{v}}_{{\mathbf{k}}} \cdot {\mathbf{\nabla }}T)^3
\frac{{\partial}^3 f_{{\mathbf{k}} }^0}{\partial T^3 }
+O\big(({\mathbf{\nabla }}T)^2({\mathbf{\nabla }}B)^1\big)
+O\big(({\mathbf{\nabla }}T)^1({\mathbf{\nabla }}B)^2\big)
+O\big(({\mathbf{\nabla }}B)^3\big),
\end{align}
and we neglect the other third-order terms
such as 
$O\big(({\mathbf{\nabla }}T)^2({\mathbf{\nabla }}B)^1\big)$,
$O\big(({\mathbf{\nabla }}T)^1({\mathbf{\nabla }}B)^2\big)$,
and
$O\big(({\mathbf{\nabla }}B)^3\big)$.
Using the relation at low temperatures
$k_{\text{B}}T  \ll  \Delta $,
\begin{align}
\frac{{\partial}^3 f_{{\mathbf{k}} }^0}{\partial T^3 }
\stackrel{\rightarrow }{=}
{\text{e}}^{-\beta \epsilon _{{{\mathbf{k}} }}}
\frac{(\beta \epsilon _{{{\mathbf{k}} }})^3}{T^3},
\end{align}
we obtain the component as
\begin{align}
g_3 \stackrel{\rightarrow }{=}
-{\tau}^3 ({\mathbf{v}}_{{\mathbf{k}}} \cdot {\mathbf{\nabla }}T)^3
{\text{e}}^{-\beta \epsilon _{{{\mathbf{k}} }}}
\frac{(\beta \epsilon _{{{\mathbf{k}} }})^3}{T^3}.
\end{align}
Assuming magnons which carry the spin angular momentum $-1$
in units of $\hbar$,
we define the spin current density
${\mathbf{j}}^{\text{s}}$
and the energy current density
${\mathbf{j}}^{\text{E}}$
as
\begin{subequations}
\begin{align}
{\mathbf{j}}^{\text{s}} =& -\int \frac{d^3{\mathbf{k}}}{(2 \pi)^3 }
g \mu_{\text{B}} {\mathbf{v}}_{{\mathbf{k}}} g_{{\mathbf{k}} },   \\
{\mathbf{j}}^{\text{E}} =& \int  \frac{d^3{\mathbf{k}}}{(2 \pi)^3 }
\epsilon _{{\mathbf{k}} }   {\mathbf{v}}_{{\mathbf{k}}} g_{{\mathbf{k}}}. 
\end{align}
\end{subequations}
Substituting 
the function
$g_{{\mathbf{k}} }=g_1+g_2+g_3$
into each current density
and performing the Gaussian integrals,
we obtain the transport coefficient
of the nonlinear response
at low temperatures
as 
\begin{subequations}
\begin{align}
 L_{13}&=L_{14}=L_{15}=L_{23}=L_{24}=L_{25}=0,  \\
 L_{16}& \stackrel{\rightarrow }{=} 
 -g \mu _{\text{B}}F {\text{e}}^{-b}
\Big[
\frac{D^3}{(\beta D)^{13/2}}A_6
+\frac{3D^2\Delta}{(\beta D)^{11/2}}A_5   
+\frac{3D{\Delta}^2}{(\beta D)^{9/2}}A_4
+\frac{{\Delta}^3}{(\beta D)^{7/2}}A_3
\Big],  \\
 L_{26}& \stackrel{\rightarrow }{=} 
F {\text{e}}^{-b}
\Big[
\frac{D^4}{(\beta D)^{15/2}}A_7
+\frac{4D^3\Delta}{(\beta D)^{13/2}}A_6  
+\frac{6D^2{\Delta}^2}{(\beta D)^{11/2}}A_5
+\frac{4D{\Delta}^3}{(\beta D)^{9/2}}A_4 
+\frac{{\Delta}^4}{(\beta D)^{7/2}}A_3
\Big],  \\
 \frac{L_{26}}{L_{16}}& \stackrel{\rightarrow }{=} 
 -\frac{\Delta}{g \mu _{\text{B}}},
\end{align}
\end{subequations}
where
$A_n:={\sqrt{\pi}}(2n)!/[2^{2n+1}(n!)] $
for 
$n \in {\mathbb{N}}$,
{$F:=({2D}/{\hbar})^4[{{\tau}^3{\beta}^3}/({{10}{\pi}^2})]$,
and $b:= \beta \Delta $}.
The second-order response vanishes
due to the property of the odd function,
and this results in 
$K_2=0$.
See Ref.~\cite{KSJD} 
for the transport coefficient of the linear response,
where 
the Onsager relation holds   
$L_{12}=L_{21}$.
At low temperatures, it reduces to~\cite{KSJD}
$  L_{12}/L_{11}=L_{21}/L_{11}
\stackrel{\rightarrow }{=}  -{\Delta}/({g \mu _{\text{B}}}) $.
Therefore,
the thermal transport coefficient of the third-order nonlinear response
$ K_3$ 
at low temperature is given as
\begin{align}
 K_3 \stackrel{\rightarrow }{=} 
\frac{1}{T^3}
\Big(L_{26}+\frac{{\Delta}}{{g \mu _{\text{B}}}}L_{16}
\Big).
\label{eqn:K3IISM}
\end{align}
Finally, substituting $ L_{26} $ and $L_{16}$
into Eq.~\eqref{eqn:K3IISM}
we obtain 
the thermal transport coefficient of the third-order nonlinear response
$K_3$
at low temperatures
in the main text.

Next,
we remark on the relaxation time.
The relaxation time 
is different from the lifetime of magnons in general.
Those are distinct quantities.
However,
under the assumption that 
impurities are dilute
and 
scattering is elastic and spatially isotopic
with the impurity potential localized in space
(i.e., the relaxation time depends solely on the magnitude of the wavenumber),
the relaxation time coincides with the lifetime.
At low temperatures
$k_{\text{B}}T  \ll  \Delta $,
the relaxation time reduces to~\cite{QBEmagnon}
\begin{align}
\tau \stackrel{\rightarrow }{=} \frac{1}{2\alpha}\frac{\hbar}{\Delta},
\end{align}
where
$\alpha$ is the Gilbert damping constant.
Since the Gilbert damping constant is little influenced by temperature~\cite{LLGspintroReview}
(i.e., the temperature dependence is negligibly small),
it is concluded that 
at low temperatures
the relaxation time takes a constant value of being temperature-independent.
Note that 
at sufficiently low temperatures,
the effect of magnon-magnon interactions and
that of phonons 
are negligibly small,
and
impurity scattering 
makes a major contribution to the relaxation.
See Ref.~\cite{QBEmagnon} for details.

Lastly, 
we comment on the Boltzmann equation.
Throughout this paper,
we study magnon transport using the conventional Boltzmann equation
of the quasiparticle approximation.
From the viewpoint of quantum field theory,
under the assumption that the variation of the center-of-mass coordinates 
is slow compared with that of the relative coordinates,
the quantum kinetic equation of the lowest order gradient approximation
becomes the quantum Boltzmann equation,
which reduces to the conventional Boltzmann equation
in the limit of the quasiparticle approximation~\cite{mahan,haug}.
Therefore, the criterion for magnon transport to be described 
by the conventional Boltzmann equation is
whether or not the quasiparticle approximation is applicable to the system.
From Ref.~\cite{QBEmagnon},
this results in the condition
$ \Delta \gg  \hbar/(2 \tau) $.
Since 
$ \hbar/(2 \tau)  \stackrel{\rightarrow }{=}  \alpha \Delta  $
at low temperatures
and 
$ \alpha \leq O(10^{-3})$
for insulating magnets~\cite{LLGspintroReview,GilbertInsulator},
the condition is satisfied.
Thus, it is concluded that 
magnon transport we study in this paper
is described by the conventional Boltzmann equation.

\end{widetext}

\bibliography{PumpingRef}

\begin{thebibliography}{55}%
\makeatletter
\providecommand \@ifxundefined [1]{%
 \@ifx{#1\undefined}
}%
\providecommand \@ifnum [1]{%
 \ifnum #1\expandafter \@firstoftwo
 \else \expandafter \@secondoftwo
 \fi
}%
\providecommand \@ifx [1]{%
 \ifx #1\expandafter \@firstoftwo
 \else \expandafter \@secondoftwo
 \fi
}%
\providecommand \natexlab [1]{#1}%
\providecommand \enquote  [1]{``#1''}%
\providecommand \bibnamefont  [1]{#1}%
\providecommand \bibfnamefont [1]{#1}%
\providecommand \citenamefont [1]{#1}%
\providecommand \href@noop [0]{\@secondoftwo}%
\providecommand \href [0]{\begingroup \@sanitize@url \@href}%
\providecommand \@href[1]{\@@startlink{#1}\@@href}%
\providecommand \@@href[1]{\endgroup#1\@@endlink}%
\providecommand \@sanitize@url [0]{\catcode `\\12\catcode `\$12\catcode
  `\&12\catcode `\#12\catcode `\^12\catcode `\_12\catcode `\%12\relax}%
\providecommand \@@startlink[1]{}%
\providecommand \@@endlink[0]{}%
\providecommand \url  [0]{\begingroup\@sanitize@url \@url }%
\providecommand \@url [1]{\endgroup\@href {#1}{\urlprefix }}%
\providecommand \urlprefix  [0]{URL }%
\providecommand \Eprint [0]{\href }%
\providecommand \doibase [0]{http://dx.doi.org/}%
\providecommand \selectlanguage [0]{\@gobble}%
\providecommand \bibinfo  [0]{\@secondoftwo}%
\providecommand \bibfield  [0]{\@secondoftwo}%
\providecommand \translation [1]{[#1]}%
\providecommand \BibitemOpen [0]{}%
\providecommand \bibitemStop [0]{}%
\providecommand \bibitemNoStop [0]{.\EOS\space}%
\providecommand \EOS [0]{\spacefactor3000\relax}%
\providecommand \BibitemShut  [1]{\csname bibitem#1\endcsname}%
\let\auto@bib@innerbib\@empty
\bibitem [{\citenamefont {Franz}\ and\ \citenamefont
  {Wiedemann}(1853)}]{WFgermany}%
  \BibitemOpen
  \bibfield  {author} {\bibinfo {author} {\bibfnamefont {R.}~\bibnamefont
  {Franz}}\ and\ \bibinfo {author} {\bibfnamefont {G.}~\bibnamefont
  {Wiedemann}},\ }\href
  {https://onlinelibrary.wiley.com/doi/abs/10.1002/andp.18531650802} {\bibfield
   {journal} {\bibinfo  {journal} {Annalen der Physik}\ }\textbf {\bibinfo
  {volume} {165}},\ \bibinfo {pages} {497} (\bibinfo {year}
  {1853})}\BibitemShut {NoStop}%
\bibitem [{\citenamefont {Lorenz}(1881)}]{Lorenz1881}%
  \BibitemOpen
  \bibfield  {author} {\bibinfo {author} {\bibfnamefont {L.}~\bibnamefont
  {Lorenz}},\ }\href
  {https://onlinelibrary.wiley.com/doi/10.1002/andp.18812490704} {\bibfield
  {journal} {\bibinfo  {journal} {Ann. Phys.}\ }\textbf {\bibinfo {volume}
  {249}},\ \bibinfo {pages} {422} (\bibinfo {year} {1881})}\BibitemShut
  {NoStop}%
\bibitem [{\citenamefont {Sommerfeld}(1928)}]{Sommerfeld1928}%
  \BibitemOpen
  \bibfield  {author} {\bibinfo {author} {\bibfnamefont {A.}~\bibnamefont
  {Sommerfeld}},\ }\href
  {https://link.springer.com/article/10.1007%2FBF01391052} {\bibfield
  {journal} {\bibinfo  {journal} {Z. Phys.}\ }\textbf {\bibinfo {volume}
  {47}},\ \bibinfo {pages} {1} (\bibinfo {year} {1928})}\BibitemShut {NoStop}%
\bibitem [{\citenamefont {Lifshitz}\ and\ \citenamefont
  {Pitaevskii}(1981)}]{LandauWF}%
  \BibitemOpen
  \bibfield  {author} {\bibinfo {author} {\bibfnamefont {E.~M.}\ \bibnamefont
  {Lifshitz}}\ and\ \bibinfo {author} {\bibfnamefont {L.~P.}\ \bibnamefont
  {Pitaevskii}},\ }\href@noop {} {\emph {\bibinfo {title} {Physical Kinetics:
  Landau and Lifshitz Course of Theoretical Physics Volume 10}}}\ (\bibinfo
  {publisher} {Pergamon Press},\ \bibinfo {year} {1981})\BibitemShut {NoStop}%
\bibitem [{\citenamefont {Ashcroft}\ and\ \citenamefont
  {Mermin}(1976)}]{AMermin}%
  \BibitemOpen
  \bibfield  {author} {\bibinfo {author} {\bibfnamefont {N.~W.}\ \bibnamefont
  {Ashcroft}}\ and\ \bibinfo {author} {\bibfnamefont {N.~D.}\ \bibnamefont
  {Mermin}},\ }\href@noop {} {\emph {\bibinfo {title} {Solid State Physics}}}\
  (\bibinfo  {publisher} {Brooks Cole, Belmont, CA},\ \bibinfo {year}
  {1976})\BibitemShut {NoStop}%
\bibitem [{\citenamefont {Chumak}\ \emph {et~al.}(2015)\citenamefont {Chumak},
  \citenamefont {Vasyuchka}, \citenamefont {Serga},\ and\ \citenamefont
  {Hillebrands}}]{MagnonSpintronics}%
  \BibitemOpen
  \bibfield  {author} {\bibinfo {author} {\bibfnamefont {A.~V.}\ \bibnamefont
  {Chumak}}, \bibinfo {author} {\bibfnamefont {V.~I.}\ \bibnamefont
  {Vasyuchka}}, \bibinfo {author} {\bibfnamefont {A.~A.}\ \bibnamefont
  {Serga}}, \ and\ \bibinfo {author} {\bibfnamefont {B.}~\bibnamefont
  {Hillebrands}},\ }\href {https://www.nature.com/articles/nphys3347}
  {\bibfield  {journal} {\bibinfo  {journal} {Nat. Phys.}\ }\textbf {\bibinfo
  {volume} {11}},\ \bibinfo {pages} {453} (\bibinfo {year} {2015})}\BibitemShut
  {NoStop}%
\bibitem [{\citenamefont {Nakata}\ \emph
  {et~al.}(2017{\natexlab{a}})\citenamefont {Nakata}, \citenamefont {Simon},\
  and\ \citenamefont {Loss}}]{ReviewMagnon}%
  \BibitemOpen
  \bibfield  {author} {\bibinfo {author} {\bibfnamefont {K.}~\bibnamefont
  {Nakata}}, \bibinfo {author} {\bibfnamefont {P.}~\bibnamefont {Simon}}, \
  and\ \bibinfo {author} {\bibfnamefont {D.}~\bibnamefont {Loss}},\ }\href
  {https://iopscience.iop.org/article/10.1088/1361-6463/aa5b09} {\bibfield
  {journal} {\bibinfo  {journal} {J. Phys. D: Appl. Phys.}\ }\textbf {\bibinfo
  {volume} {50}},\ \bibinfo {pages} {114004} (\bibinfo {year}
  {2017}{\natexlab{a}})}\BibitemShut {NoStop}%
\bibitem [{Note1()}]{Note1}%
  \BibitemOpen
  \bibinfo {note} {{We refer to the spin analog of the figure of merit for
  thermoelectric conversion elements as that for thermomagnetic ones
  $Z_{\protect \text {s}}$, i.e., $Z_{\protect \text {s}}:= {\protect \mathcal
  {S}}^2 G/K$, where ${\protect \mathcal {S}} $, $G$, and $K$ are the spin
  Seebeck coefficient, the spin conductivity, and the thermal conductivity of
  magnons, respectively.}}\BibitemShut {Stop}%
\bibitem [{\citenamefont {Nakata}\ \emph {et~al.}(2015)\citenamefont {Nakata},
  \citenamefont {Simon},\ and\ \citenamefont {Loss}}]{magnonWF}%
  \BibitemOpen
  \bibfield  {author} {\bibinfo {author} {\bibfnamefont {K.}~\bibnamefont
  {Nakata}}, \bibinfo {author} {\bibfnamefont {P.}~\bibnamefont {Simon}}, \
  and\ \bibinfo {author} {\bibfnamefont {D.}~\bibnamefont {Loss}},\ }\href
  {https://link.aps.org/doi/10.1103/PhysRevB.92.134425} {\bibfield  {journal}
  {\bibinfo  {journal} {Phys. Rev. B}\ }\textbf {\bibinfo {volume} {92}},\
  \bibinfo {pages} {134425} (\bibinfo {year} {2015})}\BibitemShut {NoStop}%
\bibitem [{\citenamefont {Nakata}\ \emph
  {et~al.}(2017{\natexlab{b}})\citenamefont {Nakata}, \citenamefont {Kim},
  \citenamefont {Klinovaja},\ and\ \citenamefont {Loss}}]{KSJD}%
  \BibitemOpen
  \bibfield  {author} {\bibinfo {author} {\bibfnamefont {K.}~\bibnamefont
  {Nakata}}, \bibinfo {author} {\bibfnamefont {S.~K.}\ \bibnamefont {Kim}},
  \bibinfo {author} {\bibfnamefont {J.}~\bibnamefont {Klinovaja}}, \ and\
  \bibinfo {author} {\bibfnamefont {D.}~\bibnamefont {Loss}},\ }\href
  {https://link.aps.org/doi/10.1103/PhysRevB.96.224414} {\bibfield  {journal}
  {\bibinfo  {journal} {Phys. Rev. B}\ }\textbf {\bibinfo {volume} {96}},\
  \bibinfo {pages} {224414} (\bibinfo {year} {2017}{\natexlab{b}})}\BibitemShut
  {NoStop}%
\bibitem [{\citenamefont {Nakata}\ \emph
  {et~al.}(2017{\natexlab{c}})\citenamefont {Nakata}, \citenamefont
  {Klinovaja},\ and\ \citenamefont {Loss}}]{KJD}%
  \BibitemOpen
  \bibfield  {author} {\bibinfo {author} {\bibfnamefont {K.}~\bibnamefont
  {Nakata}}, \bibinfo {author} {\bibfnamefont {J.}~\bibnamefont {Klinovaja}}, \
  and\ \bibinfo {author} {\bibfnamefont {D.}~\bibnamefont {Loss}},\ }\href
  {https://journals.aps.org/prb/abstract/10.1103/PhysRevB.95.125429} {\bibfield
   {journal} {\bibinfo  {journal} {Phys. Rev. B}\ }\textbf {\bibinfo {volume}
  {95}},\ \bibinfo {pages} {125429} (\bibinfo {year}
  {2017}{\natexlab{c}})}\BibitemShut {NoStop}%
\bibitem [{\citenamefont {Vera-Marun}\ \emph {et~al.}(2012)\citenamefont
  {Vera-Marun}, \citenamefont {Ranjan},\ and\ \citenamefont {van
  Wees}}]{NonlinearSpinSeebeck}%
  \BibitemOpen
  \bibfield  {author} {\bibinfo {author} {\bibfnamefont {I.~J.}\ \bibnamefont
  {Vera-Marun}}, \bibinfo {author} {\bibfnamefont {V.}~\bibnamefont {Ranjan}},
  \ and\ \bibinfo {author} {\bibfnamefont {B.~J.}\ \bibnamefont {van Wees}},\
  }\href {https://www.nature.com/articles/nphys2219#citeas} {\bibfield
  {journal} {\bibinfo  {journal} {Nat. Phys.}\ }\textbf {\bibinfo {volume}
  {8}},\ \bibinfo {pages} {313} (\bibinfo {year} {2012})}\BibitemShut {NoStop}%
\bibitem [{\citenamefont {Sakimura}\ \emph {et~al.}(2014)\citenamefont
  {Sakimura}, \citenamefont {Tashiro},\ and\ \citenamefont
  {Ando}}]{NonlinearAndo2014}%
  \BibitemOpen
  \bibfield  {author} {\bibinfo {author} {\bibfnamefont {H.}~\bibnamefont
  {Sakimura}}, \bibinfo {author} {\bibfnamefont {T.}~\bibnamefont {Tashiro}}, \
  and\ \bibinfo {author} {\bibfnamefont {K.}~\bibnamefont {Ando}},\ }\href
  {https://www.nature.com/articles/ncomms6730#citeas} {\bibfield  {journal}
  {\bibinfo  {journal} {Nat. Commun.}\ }\textbf {\bibinfo {volume} {5}},\
  \bibinfo {pages} {5730} (\bibinfo {year} {2014})}\BibitemShut {NoStop}%
\bibitem [{\citenamefont {Omar}\ \emph {et~al.}(2020)\citenamefont {Omar},
  \citenamefont {Gurram}, \citenamefont {Watanabe}, \citenamefont {Taniguchi},
  \citenamefont {Guimar\~aes},\ and\ \citenamefont {van Wees}}]{Nonlinear2020}%
  \BibitemOpen
  \bibfield  {author} {\bibinfo {author} {\bibfnamefont {S.}~\bibnamefont
  {Omar}}, \bibinfo {author} {\bibfnamefont {M.}~\bibnamefont {Gurram}},
  \bibinfo {author} {\bibfnamefont {K.}~\bibnamefont {Watanabe}}, \bibinfo
  {author} {\bibfnamefont {T.}~\bibnamefont {Taniguchi}}, \bibinfo {author}
  {\bibfnamefont {M.}~\bibnamefont {Guimar\~aes}}, \ and\ \bibinfo {author}
  {\bibfnamefont {B.}~\bibnamefont {van Wees}},\ }\href {\doibase
  10.1103/PhysRevApplied.14.064053} {\bibfield  {journal} {\bibinfo  {journal}
  {Phys. Rev. Applied}\ }\textbf {\bibinfo {volume} {14}},\ \bibinfo {pages}
  {064053} (\bibinfo {year} {2020})}\BibitemShut {NoStop}%
\bibitem [{\citenamefont {Zhang}\ \emph {et~al.}(2021)\citenamefont {Zhang},
  \citenamefont {Zhu},\ and\ \citenamefont {Su}}]{NonlinearSpinCurrent}%
  \BibitemOpen
  \bibfield  {author} {\bibinfo {author} {\bibfnamefont {Z.-F.}\ \bibnamefont
  {Zhang}}, \bibinfo {author} {\bibfnamefont {Z.-G.}\ \bibnamefont {Zhu}}, \
  and\ \bibinfo {author} {\bibfnamefont {G.}~\bibnamefont {Su}},\ }\href
  {\doibase 10.1103/PhysRevB.104.115140} {\bibfield  {journal} {\bibinfo
  {journal} {Phys. Rev. B}\ }\textbf {\bibinfo {volume} {104}},\ \bibinfo
  {pages} {115140} (\bibinfo {year} {2021})}\BibitemShut {NoStop}%
\bibitem [{\citenamefont {Tanikawa}\ \emph {et~al.}(2021)\citenamefont
  {Tanikawa}, \citenamefont {Takasan},\ and\ \citenamefont
  {Katsura}}]{KatsuraNonlinearSpinDrude}%
  \BibitemOpen
  \bibfield  {author} {\bibinfo {author} {\bibfnamefont {Y.}~\bibnamefont
  {Tanikawa}}, \bibinfo {author} {\bibfnamefont {K.}~\bibnamefont {Takasan}}, \
  and\ \bibinfo {author} {\bibfnamefont {H.}~\bibnamefont {Katsura}},\ }\href
  {\doibase 10.1103/PhysRevB.103.L201120} {\bibfield  {journal} {\bibinfo
  {journal} {Phys. Rev. B}\ }\textbf {\bibinfo {volume} {103}},\ \bibinfo
  {pages} {L201120} (\bibinfo {year} {2021})}\BibitemShut {NoStop}%
\bibitem [{\citenamefont {Nakai}\ and\ \citenamefont
  {Nagaosa}(2019)}]{NakaiNagaosa2019}%
  \BibitemOpen
  \bibfield  {author} {\bibinfo {author} {\bibfnamefont {R.}~\bibnamefont
  {Nakai}}\ and\ \bibinfo {author} {\bibfnamefont {N.}~\bibnamefont
  {Nagaosa}},\ }\href {\doibase 10.1103/PhysRevB.99.115201} {\bibfield
  {journal} {\bibinfo  {journal} {Phys. Rev. B}\ }\textbf {\bibinfo {volume}
  {99}},\ \bibinfo {pages} {115201} (\bibinfo {year} {2019})}\BibitemShut
  {NoStop}%
\bibitem [{\citenamefont {Zeng}\ \emph {et~al.}(2019)\citenamefont {Zeng},
  \citenamefont {Nandy}, \citenamefont {Taraphder},\ and\ \citenamefont
  {Tewari}}]{NonlinearNernst2019}%
  \BibitemOpen
  \bibfield  {author} {\bibinfo {author} {\bibfnamefont {C.}~\bibnamefont
  {Zeng}}, \bibinfo {author} {\bibfnamefont {S.}~\bibnamefont {Nandy}},
  \bibinfo {author} {\bibfnamefont {A.}~\bibnamefont {Taraphder}}, \ and\
  \bibinfo {author} {\bibfnamefont {S.}~\bibnamefont {Tewari}},\ }\href
  {\doibase 10.1103/PhysRevB.100.245102} {\bibfield  {journal} {\bibinfo
  {journal} {Phys. Rev. B}\ }\textbf {\bibinfo {volume} {100}},\ \bibinfo
  {pages} {245102} (\bibinfo {year} {2019})}\BibitemShut {NoStop}%
\bibitem [{\citenamefont {Zeng}\ \emph {et~al.}(2020)\citenamefont {Zeng},
  \citenamefont {Nandy},\ and\ \citenamefont {Tewari}}]{NonlinearWF2020}%
  \BibitemOpen
  \bibfield  {author} {\bibinfo {author} {\bibfnamefont {C.}~\bibnamefont
  {Zeng}}, \bibinfo {author} {\bibfnamefont {S.}~\bibnamefont {Nandy}}, \ and\
  \bibinfo {author} {\bibfnamefont {S.}~\bibnamefont {Tewari}},\ }\href
  {\doibase 10.1103/PhysRevResearch.2.032066} {\bibfield  {journal} {\bibinfo
  {journal} {Phys. Rev. Research}\ }\textbf {\bibinfo {volume} {2}},\ \bibinfo
  {pages} {032066} (\bibinfo {year} {2020})}\BibitemShut {NoStop}%
\bibitem [{\citenamefont {Watanabe}\ and\ \citenamefont
  {Oshikawa}(2020)}]{OshikawaNonlinear2020}%
  \BibitemOpen
  \bibfield  {author} {\bibinfo {author} {\bibfnamefont {H.}~\bibnamefont
  {Watanabe}}\ and\ \bibinfo {author} {\bibfnamefont {M.}~\bibnamefont
  {Oshikawa}},\ }\href {\doibase 10.1103/PhysRevB.102.165137} {\bibfield
  {journal} {\bibinfo  {journal} {Phys. Rev. B}\ }\textbf {\bibinfo {volume}
  {102}},\ \bibinfo {pages} {165137} (\bibinfo {year} {2020})}\BibitemShut
  {NoStop}%
\bibitem [{Note2()}]{Note2}%
  \BibitemOpen
  \bibinfo {note} {We refer to the magnet where Berry curvatures are zero as
  the topologically trivial magnet.}\BibitemShut {Stop}%
\bibitem [{see()}]{seeAppendices}%
  \BibitemOpen
  \href@noop {} {}\bibinfo {note} {See the Appendix for details.}\BibitemShut
  {Stop}%
\bibitem [{\citenamefont {Tokura}\ and\ \citenamefont
  {Nagaosa}(2018)}]{ReviewNonreciprocal}%
  \BibitemOpen
  \bibfield  {author} {\bibinfo {author} {\bibfnamefont {Y.}~\bibnamefont
  {Tokura}}\ and\ \bibinfo {author} {\bibfnamefont {N.}~\bibnamefont
  {Nagaosa}},\ }\href {https://www.nature.com/articles/s41467-018-05759-4}
  {\bibfield  {journal} {\bibinfo  {journal} {Nat. Commun.}\ }\textbf {\bibinfo
  {volume} {9}},\ \bibinfo {pages} {3740} (\bibinfo {year} {2018})}\BibitemShut
  {NoStop}%
\bibitem [{\citenamefont {Vlaminck}\ and\ \citenamefont
  {Bailleul}(2008)}]{DopplerScience}%
  \BibitemOpen
  \bibfield  {author} {\bibinfo {author} {\bibfnamefont {V.}~\bibnamefont
  {Vlaminck}}\ and\ \bibinfo {author} {\bibfnamefont {M.}~\bibnamefont
  {Bailleul}},\ }\href {https://www.science.org/doi/10.1126/science.1162843}
  {\bibfield  {journal} {\bibinfo  {journal} {Science}\ }\textbf {\bibinfo
  {volume} {322}},\ \bibinfo {pages} {410} (\bibinfo {year}
  {2008})}\BibitemShut {NoStop}%
\bibitem [{\citenamefont {Go}\ \emph {et~al.}(2022)\citenamefont {Go},
  \citenamefont {Lee},\ and\ \citenamefont {Kim}}]{SKKNonreciprocitySpinWaves}%
  \BibitemOpen
  \bibfield  {author} {\bibinfo {author} {\bibfnamefont {G.}~\bibnamefont
  {Go}}, \bibinfo {author} {\bibfnamefont {S.}~\bibnamefont {Lee}}, \ and\
  \bibinfo {author} {\bibfnamefont {S.~K.}\ \bibnamefont {Kim}},\ }\href
  {\doibase 10.1103/PhysRevB.105.134401} {\bibfield  {journal} {\bibinfo
  {journal} {Phys. Rev. B}\ }\textbf {\bibinfo {volume} {105}},\ \bibinfo
  {pages} {134401} (\bibinfo {year} {2022})}\BibitemShut {NoStop}%
\bibitem [{\citenamefont {Basso}\ \emph {et~al.}(2016)\citenamefont {Basso},
  \citenamefont {Ferraro},\ and\ \citenamefont {Piazzi}}]{Basso2}%
  \BibitemOpen
  \bibfield  {author} {\bibinfo {author} {\bibfnamefont {V.}~\bibnamefont
  {Basso}}, \bibinfo {author} {\bibfnamefont {E.}~\bibnamefont {Ferraro}}, \
  and\ \bibinfo {author} {\bibfnamefont {M.}~\bibnamefont {Piazzi}},\ }\href
  {https://link.aps.org/doi/10.1103/PhysRevB.94.144422} {\bibfield  {journal}
  {\bibinfo  {journal} {Phys. Rev. B}\ }\textbf {\bibinfo {volume} {94}},\
  \bibinfo {pages} {144422} (\bibinfo {year} {2016})}\BibitemShut {NoStop}%
\bibitem [{\citenamefont {Cornelissen}\ \emph
  {et~al.}(2016{\natexlab{a}})\citenamefont {Cornelissen}, \citenamefont
  {Peters}, \citenamefont {Bauer}, \citenamefont {Duine},\ and\ \citenamefont
  {van Wees}}]{MagnonChemicalWees}%
  \BibitemOpen
  \bibfield  {author} {\bibinfo {author} {\bibfnamefont {L.~J.}\ \bibnamefont
  {Cornelissen}}, \bibinfo {author} {\bibfnamefont {K.~J.~H.}\ \bibnamefont
  {Peters}}, \bibinfo {author} {\bibfnamefont {G.~E.~W.}\ \bibnamefont
  {Bauer}}, \bibinfo {author} {\bibfnamefont {R.~A.}\ \bibnamefont {Duine}}, \
  and\ \bibinfo {author} {\bibfnamefont {B.~J.}\ \bibnamefont {van Wees}},\
  }\href {\doibase 10.1103/PhysRevB.94.014412} {\bibfield  {journal} {\bibinfo
  {journal} {Phys. Rev. B}\ }\textbf {\bibinfo {volume} {94}},\ \bibinfo
  {pages} {014412} (\bibinfo {year} {2016}{\natexlab{a}})}\BibitemShut
  {NoStop}%
\bibitem [{\citenamefont {Du}\ \emph {et~al.}(2017)\citenamefont {Du},
  \citenamefont {der Sar}, \citenamefont {Zhou}, \citenamefont {Upadhyaya},
  \citenamefont {Casola}, \citenamefont {Zhang}, \citenamefont {Onbasli},
  \citenamefont {Ross}, \citenamefont {Walsworth}, \citenamefont
  {Tserkovnyak},\ and\ \citenamefont {Yacoby}}]{YacobyChemical}%
  \BibitemOpen
  \bibfield  {author} {\bibinfo {author} {\bibfnamefont {C.}~\bibnamefont
  {Du}}, \bibinfo {author} {\bibfnamefont {T.~V.}\ \bibnamefont {der Sar}},
  \bibinfo {author} {\bibfnamefont {T.~X.}\ \bibnamefont {Zhou}}, \bibinfo
  {author} {\bibfnamefont {P.}~\bibnamefont {Upadhyaya}}, \bibinfo {author}
  {\bibfnamefont {F.}~\bibnamefont {Casola}}, \bibinfo {author} {\bibfnamefont
  {H.}~\bibnamefont {Zhang}}, \bibinfo {author} {\bibfnamefont {M.~C.}\
  \bibnamefont {Onbasli}}, \bibinfo {author} {\bibfnamefont {C.~A.}\
  \bibnamefont {Ross}}, \bibinfo {author} {\bibfnamefont {R.~L.}\ \bibnamefont
  {Walsworth}}, \bibinfo {author} {\bibfnamefont {Y.}~\bibnamefont
  {Tserkovnyak}}, \ and\ \bibinfo {author} {\bibfnamefont {A.}~\bibnamefont
  {Yacoby}},\ }\href {https://science.sciencemag.org/content/357/6347/195}
  {\bibfield  {journal} {\bibinfo  {journal} {Science}\ }\textbf {\bibinfo
  {volume} {357}},\ \bibinfo {pages} {195} (\bibinfo {year}
  {2017})}\BibitemShut {NoStop}%
\bibitem [{\citenamefont {Demokritov}\ \emph {et~al.}(2006)\citenamefont
  {Demokritov}, \citenamefont {Demidov}, \citenamefont {Dzyapko}, \citenamefont
  {Melkov}, \citenamefont {Serga}, \citenamefont {Hillebrands},\ and\
  \citenamefont {Slavin}}]{demokritov}%
  \BibitemOpen
  \bibfield  {author} {\bibinfo {author} {\bibfnamefont {S.~O.}\ \bibnamefont
  {Demokritov}}, \bibinfo {author} {\bibfnamefont {V.~E.}\ \bibnamefont
  {Demidov}}, \bibinfo {author} {\bibfnamefont {O.}~\bibnamefont {Dzyapko}},
  \bibinfo {author} {\bibfnamefont {G.~A.}\ \bibnamefont {Melkov}}, \bibinfo
  {author} {\bibfnamefont {A.~A.}\ \bibnamefont {Serga}}, \bibinfo {author}
  {\bibfnamefont {B.}~\bibnamefont {Hillebrands}}, \ and\ \bibinfo {author}
  {\bibfnamefont {A.~N.}\ \bibnamefont {Slavin}},\ }\href
  {https://www.nature.com/articles/nature05117} {\bibfield  {journal} {\bibinfo
   {journal} {Nature (London)}\ }\textbf {\bibinfo {volume} {443}},\ \bibinfo
  {pages} {430} (\bibinfo {year} {2006})}\BibitemShut {NoStop}%
\bibitem [{\citenamefont {Haldane}\ and\ \citenamefont
  {Arovas}(1995)}]{Haldane2}%
  \BibitemOpen
  \bibfield  {author} {\bibinfo {author} {\bibfnamefont {F.~D.~M.}\
  \bibnamefont {Haldane}}\ and\ \bibinfo {author} {\bibfnamefont {D.~P.}\
  \bibnamefont {Arovas}},\ }\href {\doibase 10.1103/PhysRevB.52.4223}
  {\bibfield  {journal} {\bibinfo  {journal} {Phys. Rev. B}\ }\textbf {\bibinfo
  {volume} {52}},\ \bibinfo {pages} {4223} (\bibinfo {year}
  {1995})}\BibitemShut {NoStop}%
\bibitem [{\citenamefont {Meier}\ and\ \citenamefont {Loss}(2003)}]{magnon2}%
  \BibitemOpen
  \bibfield  {author} {\bibinfo {author} {\bibfnamefont {F.}~\bibnamefont
  {Meier}}\ and\ \bibinfo {author} {\bibfnamefont {D.}~\bibnamefont {Loss}},\
  }\href {\doibase 10.1103/PhysRevLett.90.167204} {\bibfield  {journal}
  {\bibinfo  {journal} {Phys. Rev. Lett.}\ }\textbf {\bibinfo {volume} {90}},\
  \bibinfo {pages} {167204} (\bibinfo {year} {2003})}\BibitemShut {NoStop}%
\bibitem [{\citenamefont {Fujimoto}(2009)}]{Fujimoto}%
  \BibitemOpen
  \bibfield  {author} {\bibinfo {author} {\bibfnamefont {S.}~\bibnamefont
  {Fujimoto}},\ }\href {\doibase 10.1103/PhysRevLett.103.047203} {\bibfield
  {journal} {\bibinfo  {journal} {Phys. Rev. Lett.}\ }\textbf {\bibinfo
  {volume} {103}},\ \bibinfo {pages} {047203} (\bibinfo {year}
  {2009})}\BibitemShut {NoStop}%
\bibitem [{\citenamefont {Nakata}\ \emph {et~al.}(2018)\citenamefont {Nakata},
  \citenamefont {Ohnuma},\ and\ \citenamefont
  {Matsuo}}]{KNmagnonNoiseJunction}%
  \BibitemOpen
  \bibfield  {author} {\bibinfo {author} {\bibfnamefont {K.}~\bibnamefont
  {Nakata}}, \bibinfo {author} {\bibfnamefont {Y.}~\bibnamefont {Ohnuma}}, \
  and\ \bibinfo {author} {\bibfnamefont {M.}~\bibnamefont {Matsuo}},\ }\href
  {https://link.aps.org/doi/10.1103/PhysRevB.98.094430} {\bibfield  {journal}
  {\bibinfo  {journal} {Phys. Rev. B}\ }\textbf {\bibinfo {volume} {98}},\
  \bibinfo {pages} {094430} (\bibinfo {year} {2018})}\BibitemShut {NoStop}%
\bibitem [{Note3()}]{Note3}%
  \BibitemOpen
  \bibinfo {note} {This result changes in general if one assumes a magnon
  energy dispersion with the $k$-linear term.}\BibitemShut {Stop}%
\bibitem [{\citenamefont {Nakata}\ and\ \citenamefont
  {Ohnuma}(2021)}]{QBEmagnon}%
  \BibitemOpen
  \bibfield  {author} {\bibinfo {author} {\bibfnamefont {K.}~\bibnamefont
  {Nakata}}\ and\ \bibinfo {author} {\bibfnamefont {Y.}~\bibnamefont
  {Ohnuma}},\ }\href {\doibase 10.1103/PhysRevB.104.064408} {\bibfield
  {journal} {\bibinfo  {journal} {Phys. Rev. B}\ }\textbf {\bibinfo {volume}
  {104}},\ \bibinfo {pages} {064408} (\bibinfo {year} {2021})}\BibitemShut
  {NoStop}%
\bibitem [{\citenamefont {Tserkovnyak}\ \emph {et~al.}(2005)\citenamefont
  {Tserkovnyak}, \citenamefont {Brataas}, \citenamefont {Bauer},\ and\
  \citenamefont {Halperin}}]{LLGspintroReview}%
  \BibitemOpen
  \bibfield  {author} {\bibinfo {author} {\bibfnamefont {Y.}~\bibnamefont
  {Tserkovnyak}}, \bibinfo {author} {\bibfnamefont {A.}~\bibnamefont
  {Brataas}}, \bibinfo {author} {\bibfnamefont {G.~E.~W.}\ \bibnamefont
  {Bauer}}, \ and\ \bibinfo {author} {\bibfnamefont {B.~I.}\ \bibnamefont
  {Halperin}},\ }\href {\doibase 10.1103/RevModPhys.77.1375} {\bibfield
  {journal} {\bibinfo  {journal} {Rev. Mod. Phys.}\ }\textbf {\bibinfo {volume}
  {77}},\ \bibinfo {pages} {1375} (\bibinfo {year} {2005})}\BibitemShut
  {NoStop}%
\bibitem [{\citenamefont {S\'anchez}\ and\ \citenamefont
  {L\'opez}(2013)}]{NLelWFviolation}%
  \BibitemOpen
  \bibfield  {author} {\bibinfo {author} {\bibfnamefont {D.}~\bibnamefont
  {S\'anchez}}\ and\ \bibinfo {author} {\bibfnamefont {R.}~\bibnamefont
  {L\'opez}},\ }\href {\doibase 10.1103/PhysRevLett.110.026804} {\bibfield
  {journal} {\bibinfo  {journal} {Phys. Rev. Lett.}\ }\textbf {\bibinfo
  {volume} {110}},\ \bibinfo {pages} {026804} (\bibinfo {year}
  {2013})}\BibitemShut {NoStop}%
\bibitem [{\citenamefont {L\'opez}\ and\ \citenamefont
  {S\'anchez}(2013)}]{ViolationWF_QD}%
  \BibitemOpen
  \bibfield  {author} {\bibinfo {author} {\bibfnamefont {R.}~\bibnamefont
  {L\'opez}}\ and\ \bibinfo {author} {\bibfnamefont {D.}~\bibnamefont
  {S\'anchez}},\ }\href {\doibase 10.1103/PhysRevB.88.045129} {\bibfield
  {journal} {\bibinfo  {journal} {Phys. Rev. B}\ }\textbf {\bibinfo {volume}
  {88}},\ \bibinfo {pages} {045129} (\bibinfo {year} {2013})}\BibitemShut
  {NoStop}%
\bibitem [{Note4()}]{Note4}%
  \BibitemOpen
  \bibinfo {note} {See Ref.~\cite {kondoNonlinear} for topological magnon
  systems.}\BibitemShut {Stop}%
\bibitem [{\citenamefont {Prasai}\ \emph {et~al.}(2017)\citenamefont {Prasai},
  \citenamefont {Trump}, \citenamefont {Marcus}, \citenamefont {Akopyan},
  \citenamefont {Huang}, \citenamefont {McQueen},\ and\ \citenamefont
  {Cohn}}]{PhononMagnonPrasai}%
  \BibitemOpen
  \bibfield  {author} {\bibinfo {author} {\bibfnamefont {N.}~\bibnamefont
  {Prasai}}, \bibinfo {author} {\bibfnamefont {B.~A.}\ \bibnamefont {Trump}},
  \bibinfo {author} {\bibfnamefont {G.~G.}\ \bibnamefont {Marcus}}, \bibinfo
  {author} {\bibfnamefont {A.}~\bibnamefont {Akopyan}}, \bibinfo {author}
  {\bibfnamefont {S.~X.}\ \bibnamefont {Huang}}, \bibinfo {author}
  {\bibfnamefont {T.~M.}\ \bibnamefont {McQueen}}, \ and\ \bibinfo {author}
  {\bibfnamefont {J.~L.}\ \bibnamefont {Cohn}},\ }\href {\doibase
  10.1103/PhysRevB.95.224407} {\bibfield  {journal} {\bibinfo  {journal} {Phys.
  Rev. B}\ }\textbf {\bibinfo {volume} {95}},\ \bibinfo {pages} {224407}
  (\bibinfo {year} {2017})}\BibitemShut {NoStop}%
\bibitem [{\citenamefont {Artman}\ \emph {et~al.}(1965)\citenamefont {Artman},
  \citenamefont {Murphy},\ and\ \citenamefont {Foner}}]{CrOexp}%
  \BibitemOpen
  \bibfield  {author} {\bibinfo {author} {\bibfnamefont {J.~O.}\ \bibnamefont
  {Artman}}, \bibinfo {author} {\bibfnamefont {J.~C.}\ \bibnamefont {Murphy}},
  \ and\ \bibinfo {author} {\bibfnamefont {S.}~\bibnamefont {Foner}},\ }\href
  {\doibase 10.1103/PhysRev.138.A912} {\bibfield  {journal} {\bibinfo
  {journal} {Phys. Rev.}\ }\textbf {\bibinfo {volume} {138}},\ \bibinfo {pages}
  {A912} (\bibinfo {year} {1965})}\BibitemShut {NoStop}%
\bibitem [{\citenamefont {Kota}\ and\ \citenamefont {Imamura}(2017)}]{CrOexp2}%
  \BibitemOpen
  \bibfield  {author} {\bibinfo {author} {\bibfnamefont {Y.}~\bibnamefont
  {Kota}}\ and\ \bibinfo {author} {\bibfnamefont {H.}~\bibnamefont {Imamura}},\
  }\href {https://iopscience.iop.org/article/10.7567/APEX.10.013002} {\bibfield
   {journal} {\bibinfo  {journal} {Appl. Phys. Express}\ }\textbf {\bibinfo
  {volume} {10}},\ \bibinfo {pages} {013002} (\bibinfo {year}
  {2017})}\BibitemShut {NoStop}%
\bibitem [{\citenamefont {Heinrich}\ \emph {et~al.}(2011)\citenamefont
  {Heinrich}, \citenamefont {Burrowes}, \citenamefont {Montoya}, \citenamefont
  {Kardasz}, \citenamefont {Girt}, \citenamefont {Song}, \citenamefont {Sun},\
  and\ \citenamefont {Wu}}]{GilbertInsulator}%
  \BibitemOpen
  \bibfield  {author} {\bibinfo {author} {\bibfnamefont {B.}~\bibnamefont
  {Heinrich}}, \bibinfo {author} {\bibfnamefont {C.}~\bibnamefont {Burrowes}},
  \bibinfo {author} {\bibfnamefont {E.}~\bibnamefont {Montoya}}, \bibinfo
  {author} {\bibfnamefont {B.}~\bibnamefont {Kardasz}}, \bibinfo {author}
  {\bibfnamefont {E.}~\bibnamefont {Girt}}, \bibinfo {author} {\bibfnamefont
  {Y.-Y.}\ \bibnamefont {Song}}, \bibinfo {author} {\bibfnamefont
  {Y.}~\bibnamefont {Sun}}, \ and\ \bibinfo {author} {\bibfnamefont
  {M.}~\bibnamefont {Wu}},\ }\href {\doibase 10.1103/PhysRevLett.107.066604}
  {\bibfield  {journal} {\bibinfo  {journal} {Phys. Rev. Lett.}\ }\textbf
  {\bibinfo {volume} {107}},\ \bibinfo {pages} {066604} (\bibinfo {year}
  {2011})}\BibitemShut {NoStop}%
\bibitem [{\citenamefont {Chudo}()}]{ChudoPrivate}%
  \BibitemOpen
  \bibfield  {author} {\bibinfo {author} {\bibfnamefont {H.}~\bibnamefont
  {Chudo}},\ }\href@noop {} {}\bibinfo {note} {Private
  communication}\BibitemShut {NoStop}%
\bibitem [{\citenamefont {Cornelissen}\ \emph {et~al.}(2015)\citenamefont
  {Cornelissen}, \citenamefont {Liu}, \citenamefont {Duine}, \citenamefont
  {Youssef},\ and\ \citenamefont {van Wees}}]{WeesNatPhys}%
  \BibitemOpen
  \bibfield  {author} {\bibinfo {author} {\bibfnamefont {L.~J.}\ \bibnamefont
  {Cornelissen}}, \bibinfo {author} {\bibfnamefont {J.}~\bibnamefont {Liu}},
  \bibinfo {author} {\bibfnamefont {R.~A.}\ \bibnamefont {Duine}}, \bibinfo
  {author} {\bibfnamefont {J.~B.}\ \bibnamefont {Youssef}}, \ and\ \bibinfo
  {author} {\bibfnamefont {B.~J.}\ \bibnamefont {van Wees}},\ }\href
  {https://www.nature.com/articles/nphys3465} {\bibfield  {journal} {\bibinfo
  {journal} {Nat. Phys.}\ }\textbf {\bibinfo {volume} {11}},\ \bibinfo {pages}
  {1022} (\bibinfo {year} {2015})}\BibitemShut {NoStop}%
\bibitem [{\citenamefont {Cornelissen}\ \emph
  {et~al.}(2016{\natexlab{b}})\citenamefont {Cornelissen}, \citenamefont
  {Shan},\ and\ \citenamefont {van Wees}}]{MagnonG}%
  \BibitemOpen
  \bibfield  {author} {\bibinfo {author} {\bibfnamefont {L.~J.}\ \bibnamefont
  {Cornelissen}}, \bibinfo {author} {\bibfnamefont {J.}~\bibnamefont {Shan}}, \
  and\ \bibinfo {author} {\bibfnamefont {B.~J.}\ \bibnamefont {van Wees}},\
  }\href {\doibase 10.1103/PhysRevB.94.180402} {\bibfield  {journal} {\bibinfo
  {journal} {Phys. Rev. B}\ }\textbf {\bibinfo {volume} {94}},\ \bibinfo
  {pages} {180402} (\bibinfo {year} {2016}{\natexlab{b}})}\BibitemShut
  {NoStop}%
\bibitem [{\citenamefont {Onose}\ \emph {et~al.}(2010)\citenamefont {Onose},
  \citenamefont {Ideue}, \citenamefont {Katsura}, \citenamefont {Shiomi},
  \citenamefont {Nagaosa},\ and\ \citenamefont {Tokura}}]{onose}%
  \BibitemOpen
  \bibfield  {author} {\bibinfo {author} {\bibfnamefont {Y.}~\bibnamefont
  {Onose}}, \bibinfo {author} {\bibfnamefont {T.}~\bibnamefont {Ideue}},
  \bibinfo {author} {\bibfnamefont {H.}~\bibnamefont {Katsura}}, \bibinfo
  {author} {\bibfnamefont {Y.}~\bibnamefont {Shiomi}}, \bibinfo {author}
  {\bibfnamefont {N.}~\bibnamefont {Nagaosa}}, \ and\ \bibinfo {author}
  {\bibfnamefont {Y.}~\bibnamefont {Tokura}},\ }\href
  {https://science.sciencemag.org/content/329/5989/297.editor-summary}
  {\bibfield  {journal} {\bibinfo  {journal} {Science}\ }\textbf {\bibinfo
  {volume} {329}},\ \bibinfo {pages} {297} (\bibinfo {year}
  {2010})}\BibitemShut {NoStop}%
\bibitem [{\citenamefont {Tabuchi}\ \emph {et~al.}(2014)\citenamefont
  {Tabuchi}, \citenamefont {Ichino}, \citenamefont {Ishikawa}, \citenamefont
  {Yamazaki}, \citenamefont {Usami},\ and\ \citenamefont {Nakamura}}]{tabuchi}%
  \BibitemOpen
  \bibfield  {author} {\bibinfo {author} {\bibfnamefont {Y.}~\bibnamefont
  {Tabuchi}}, \bibinfo {author} {\bibfnamefont {S.}~\bibnamefont {Ichino}},
  \bibinfo {author} {\bibfnamefont {T.}~\bibnamefont {Ishikawa}}, \bibinfo
  {author} {\bibfnamefont {R.}~\bibnamefont {Yamazaki}}, \bibinfo {author}
  {\bibfnamefont {K.}~\bibnamefont {Usami}}, \ and\ \bibinfo {author}
  {\bibfnamefont {Y.}~\bibnamefont {Nakamura}},\ }\href
  {https://journals.aps.org/prl/abstract/10.1103/PhysRevLett.113.083603}
  {\bibfield  {journal} {\bibinfo  {journal} {Phys. Rev. Lett.}\ }\textbf
  {\bibinfo {volume} {113}},\ \bibinfo {pages} {083603} (\bibinfo {year}
  {2014})}\BibitemShut {NoStop}%
\bibitem [{\citenamefont {Tabuchi}\ \emph {et~al.}(2015)\citenamefont
  {Tabuchi}, \citenamefont {Ichino}, \citenamefont {Noguchi}, \citenamefont
  {Ishikawa}, \citenamefont {Yamazaki}, \citenamefont {Usami},\ and\
  \citenamefont {Nakamura}}]{tabuchiScience}%
  \BibitemOpen
  \bibfield  {author} {\bibinfo {author} {\bibfnamefont {Y.}~\bibnamefont
  {Tabuchi}}, \bibinfo {author} {\bibfnamefont {S.}~\bibnamefont {Ichino}},
  \bibinfo {author} {\bibfnamefont {A.}~\bibnamefont {Noguchi}}, \bibinfo
  {author} {\bibfnamefont {T.}~\bibnamefont {Ishikawa}}, \bibinfo {author}
  {\bibfnamefont {R.}~\bibnamefont {Yamazaki}}, \bibinfo {author}
  {\bibfnamefont {K.}~\bibnamefont {Usami}}, \ and\ \bibinfo {author}
  {\bibfnamefont {Y.}~\bibnamefont {Nakamura}},\ }\href
  {https://science.sciencemag.org/content/349/6246/405} {\bibfield  {journal}
  {\bibinfo  {journal} {Science}\ }\textbf {\bibinfo {volume} {349}},\ \bibinfo
  {pages} {405} (\bibinfo {year} {2015})}\BibitemShut {NoStop}%
\bibitem [{\citenamefont {Kosen}\ \emph {et~al.}(2018)\citenamefont {Kosen},
  \citenamefont {Morris}, \citenamefont {van Loo},\ and\ \citenamefont
  {Karenowska}}]{magnon10mK}%
  \BibitemOpen
  \bibfield  {author} {\bibinfo {author} {\bibfnamefont {S.}~\bibnamefont
  {Kosen}}, \bibinfo {author} {\bibfnamefont {R.~G.~E.}\ \bibnamefont
  {Morris}}, \bibinfo {author} {\bibfnamefont {A.~F.}\ \bibnamefont {van Loo}},
  \ and\ \bibinfo {author} {\bibfnamefont {A.~D.}\ \bibnamefont {Karenowska}},\
  }\href {https://aip.scitation.org/doi/10.1063/1.5011767} {\bibfield
  {journal} {\bibinfo  {journal} {Appl. Phys. Lett.}\ }\textbf {\bibinfo
  {volume} {112}},\ \bibinfo {pages} {012402} (\bibinfo {year}
  {2018})}\BibitemShut {NoStop}%
\bibitem [{\citenamefont {Mahan}(2000)}]{mahan}%
  \BibitemOpen
  \bibfield  {author} {\bibinfo {author} {\bibfnamefont {G.~D.}\ \bibnamefont
  {Mahan}},\ }\href@noop {} {\emph {\bibinfo {title} {Many-Particle Physics}}}\
  (\bibinfo  {publisher} {Kluwer Academic, Plenum Publishers, New York},\
  \bibinfo {year} {2000})\BibitemShut {NoStop}%
\bibitem [{\citenamefont {Haug}\ and\ \citenamefont {Jauho}(2007)}]{haug}%
  \BibitemOpen
  \bibfield  {author} {\bibinfo {author} {\bibfnamefont {H.}~\bibnamefont
  {Haug}}\ and\ \bibinfo {author} {\bibfnamefont {A.}~\bibnamefont {Jauho}},\
  }\href@noop {} {\emph {\bibinfo {title} {Quantum Kinetics in Transport and
  Optics of Semiconductors}}}\ (\bibinfo  {publisher} {Springer New York},\
  \bibinfo {year} {2007})\BibitemShut {NoStop}%
\bibitem [{\citenamefont {Itoh}\ \emph {et~al.}(2013)\citenamefont {Itoh},
  \citenamefont {Endoh}, \citenamefont {Yokoo}, \citenamefont {Kawana},
  \citenamefont {Kaneko}, \citenamefont {Tokura},\ and\ \citenamefont
  {Fujita}}]{SrRuO3}%
  \BibitemOpen
  \bibfield  {author} {\bibinfo {author} {\bibfnamefont {S.}~\bibnamefont
  {Itoh}}, \bibinfo {author} {\bibfnamefont {Y.}~\bibnamefont {Endoh}},
  \bibinfo {author} {\bibfnamefont {T.}~\bibnamefont {Yokoo}}, \bibinfo
  {author} {\bibfnamefont {D.}~\bibnamefont {Kawana}}, \bibinfo {author}
  {\bibfnamefont {Y.}~\bibnamefont {Kaneko}}, \bibinfo {author} {\bibfnamefont
  {Y.}~\bibnamefont {Tokura}}, \ and\ \bibinfo {author} {\bibfnamefont
  {M.}~\bibnamefont {Fujita}},\ }\href
  {https://journals.jps.jp/doi/10.7566/JPSJ.82.043001} {\bibfield  {journal}
  {\bibinfo  {journal} {Journal of the Physical Society of Japan}\ }\textbf
  {\bibinfo {volume} {82}},\ \bibinfo {pages} {043001} (\bibinfo {year}
  {2013})}\BibitemShut {NoStop}%
\bibitem [{\citenamefont {Lado}\ and\ \citenamefont
  {Fern{\'a}ndez-Rossier}(2017)}]{CrI3}%
  \BibitemOpen
  \bibfield  {author} {\bibinfo {author} {\bibfnamefont {J.~L.}\ \bibnamefont
  {Lado}}\ and\ \bibinfo {author} {\bibfnamefont {J.}~\bibnamefont
  {Fern{\'a}ndez-Rossier}},\ }\href
  {https://iopscience.iop.org/article/10.1088/2053-1583/aa75ed} {\bibfield
  {journal} {\bibinfo  {journal} {2D Materials}\ }\textbf {\bibinfo {volume}
  {4}},\ \bibinfo {pages} {035002} (\bibinfo {year} {2017})}\BibitemShut
  {NoStop}%
\bibitem [{\citenamefont {Kondo}\ and\ \citenamefont
  {Akagi}(2022)}]{kondoNonlinear}%
  \BibitemOpen
  \bibfield  {author} {\bibinfo {author} {\bibfnamefont {H.}~\bibnamefont
  {Kondo}}\ and\ \bibinfo {author} {\bibfnamefont {Y.}~\bibnamefont {Akagi}},\
  }\href {\doibase 10.1103/PhysRevResearch.4.013186} {\bibfield  {journal}
  {\bibinfo  {journal} {Phys. Rev. Research}\ }\textbf {\bibinfo {volume}
  {4}},\ \bibinfo {pages} {013186} (\bibinfo {year} {2022})}\BibitemShut
  {NoStop}%
\end{thebibliography}%

\end{document}